\documentstyle[aps,prd,preprint,tighten]{revtex}
\input epsf

\def\etal{{\it et al.}}
\def\Bg{B_\gamma}
\def\mnu{m_\nu}
\def\Tnu{T_\nu}
\def\mt{\mnu\tau}
\def\keV {\;{\rm \hbox{ke\kern-0.14em V}}}
\def\MeV {\;{\rm \hbox{Me\kern-0.14em V}}}
\def\sec {\;{\rm sec}}
\def\ie {{\it i.e.}}
\def\brem{bremsstrahlung\ }
\begin{document}
\preprint{CITA-95-26}

\title{Gamma Rays and the Decay of Neutrinos from SN1987A}

\author{Andrew H. Jaffe}
\address{Canadian Institute for Theoretical Astrophysics,\\
60 St. George St., Toronto, Ontario M5S 1A1, Canada}
\author{Michael S. Turner}
\address{
Departments of Physics and of Astronomy \& Astrophysics,\\
Enrico Fermi Institute\\
The University of Chicago, Chicago, IL 60637-1433\\
and\\
Theoretical Astrophysics, Fermi National Accelerator Laboratory\\
Batavia, IL 60510-0500}
\date{\today}

\maketitle
\begin{abstract}
  We calculate limits to the properties of massive, unstable neutrinos
  using data from gamma-ray detectors on the Pioneer Venus Orbiter
  Satellite; a massive neutrino emitted from SN1987A that decayed in
  flight and produced gamma rays would be detectable by this
  instruments. The lack of such a signal allows us to constrain the
  branching ratio to photons ($\Bg$), mass ($\mnu$), and radiative
  lifetime ($\tau_\gamma = \tau/\Bg$).  For low mass ($m<T\sim8\MeV$)
  neutrinos decaying $\nu\rightarrow\nu'\gamma$, $\Bg<3\times
  10^{-7}$, for $\mt\lesssim 10^{6} \keV\sec$, and $\Bg<6\times 10^{-14}
  \mt/\keV\sec$ for $\mt\gtrsim 10^{6} \keV\sec$; limits for high-mass
  neutrinos are somewhat weaker due to Boltzmann suppression. We also
  calculate limits for decays that produce gamma rays through the \brem
  channel, $\nu\rightarrow\nu'e^+e^-\gamma$. In the case that
  neutrino mass states are nearly degenerate, $\delta m^2/m^2\ll1$, our
  limits for the mode $\nu\rightarrow\nu'\gamma$ become more stringent
  by a factor of $\delta m^2/m^2$, because more of the decay photons are
  shifted into the PVO detector energy window.
\end{abstract}

\section{Introduction}

The occurrence of Supernova 1987A in the Large Magellanic Cloud has
proven to be among the most fruitful experiments in the heavenly
laboratory for confirming ``known'' physics and 
constraining new physics. Aside from its obvious impact upon the
study of the late stages of stellar evolution in general and upon
supernova physics in particular, models for SN1987A have become a test
bed for the study of the couplings of light particles (neutrinos,
axions) to ordinary matter\cite{ellis&schramm}.  In this work, we
discuss limits on the properties of neutrinos independent of a specific
model for the supernova based upon the emission of thermal neutrinos
from the hot nascent neutron star.

When a supernova occurs, the bulk of the binding energy of the
progenitor star ($\sim3\times10^{53}$~erg) is released in neutrinos, as
predicted by theory and confirmed by the observation of a neutrino
burst from SN1987A, with a characteristic temperature of about
$T_\nu\approx4.5$ MeV (for low-mass electron neutrinos; other species
are predicted to have a higher temperature, $T\simeq8\MeV$, because of
their different coupling to the prevalent electrons
\cite{sigl,burrowsannrev}). If at least one species of neutrinos is
unstable and couples to the photon, then some of these neutrinos will
decay to photons en route, which are potentially detectable as MeV gamma
rays. At the time of the supernova burst's arrival at earth and
environs, there were several satellites operating in the solar system
capable of detecting the decay photons in the course of their watch for
gamma-ray bursts. Analyses of the data from one of these detectors, on
board the Solar Max Mission (SMM) satellite, has already been
presented\cite{smm1}. Here, we examine the data from the Gamma Burst
Detector on the Pioneer Venus Orbiter (PVO). While the PVO detector was
smaller, and its energy window is not well-matched to that of the
supernova neutrinos, it had $4\pi$ acceptance and was in an environment
free of the Earth's radiation belts (leading to lower backgrounds). In
addition, more high quality data is available ($>8000$~sec vs. 10~sec
for SMM).

The paper is organized as follows. In the next section an exact formula
for the expected gamma-ray flux is derived and important approximations
are developed. In Section~\ref{sec:stat}, the PVO data are discussed and
rigorous limits are derived in the simplest regime. The next four
sections build upon these results, expanding to more complicated
regimes. The final section is a brief summary.

\section{Gamma-ray signal}

We can write the expected fluence of gamma rays from decaying
neutrinos with mass $\mnu$ and mean lifetime $\tau$
as\cite{procs}
\begin{eqnarray}
dN &= &{\Bg L_\#(E)\over4\pi D^2} dE \,\, n(\mu) d\mu\;
      {e^{-t_d/\gamma\tau}\over\gamma\tau} dt_d \nonumber\\ 
      &&\times\delta\left(t-t_d\left[1-v\mu + {D\over t_d}\left\{
      \sqrt{1-\left(vt_d/D\right)^2(1-\mu^2)}-1\right\}
         \right]\right)dt.
\end{eqnarray}
where $\Bg$ is the fraction of decays that produce a gamma ray.  The
first factor is the overall flux of neutrinos from a supernova at a
distance $D$.  $L_\#(E)$
is the differential number flux of neutrinos of energy $E$, so $E_T=\int
dE\;EL_\#$ is the total luminosity in neutrinos.  The second factor
gives the fraction that decay into a ``lab-frame'' angle $\arccos\mu$.
The third factor gives the fraction that decay at time $t_d$.
Finally, the delta function selects the photons with a given $t_d$, $E$,
$\mu$ that arrive at a time $t$ after the  arrival of massless neutrinos at
the detector. The Lorentz factor is $\gamma=E/\mnu$ and the
speed $v=\sqrt{1-\gamma^{-2}}$.

The function $n(\mu)$ depends on the distribution of daughter photons in
the neutrino rest frame, and therefore on the particular decay channel
involved. First, we consider the two-body decay,
$\nu\rightarrow\nu'\gamma$.  Because the neutrino is a spin-1/2
particle, and the photon a spin-1 particle, this reaction can proceed in
one of two ways: with the helicity of the daughter neutrino either
parallel or antiparallel the initial helicity.  From quantum mechanics,
then, the distribution of the photon in the rest frame of the parent
will be proportional to either $(1\pm\bar\mu)/2$, where $\bar\mu$ is the
cosine of the rest-frame angle between the directions of the parent
neutrino and the photon.  Transforming into the lab frame gives us the
distribution $n(\mu)$.  Note that we have assumed the neutrinos are
ejected from the supernova in an instantaneous burst; as long as the
actual duration of the pulse is small compared to the timing resolution
of the detector, which is the case, this is a good approximation.

Because we are not interested in the decay angle, but rather the
photon energy, we write $n(\mu)d\mu=f(E,k)dk$, where $k$ is the
gamma-ray energy, related to the decay angle by
\begin{equation}
  \mu={\gamma\over\sqrt{\gamma^2-1}}\left(1-{m\over2\gamma k}\right). 
\end{equation}
This gives
\begin{equation}
  f(E,k)={1\over(Ev)^2}\left(Ev\mp E\pm 2k\right),
\end{equation}
for each of the helicity possibilities. (For reference, an
isotropic decay would give $f(E,k)=1/(Ev)=1/p$, where $p$ is the
neutrino momentum.) For low-mass neutrinos with $v\approx1$,
\begin{equation}
  f(E,k) = \cases{2k/E^2     &no flip\cr
                  2(E-k)/E^2 &flip\cr
                  1/E        &isotropic\cr
                  }.
\end{equation}
Each of these should be multiplied by a Heaviside
function $\Theta(E-k)$ to require that the daughter photon be
less energetic than the parent. Further, we require that the
decay does not occur inside of the progenitor envelope which
would considerably alter the energetics of the explosion and lead to an
independent constraint which is important for short
lifetimes\cite{falk&schramm}.

The factor inside the delta function is especially complicated.  This
occurs because at any given time, the detector is receiving photons from
neutrinos that have decayed in a complicated shape, approximately an
ellipsoid with the supernova at one focus and the detector at another,
further complicated by the speed $v<1$ of the massive neutrinos. This
includes photons that have left the supernova pointing far away from the
detector but have decayed at large angles toward the detector.
Obviously, for low mass neutrinos which leave the supernova at highly
relativistic speeds, the fraction which take such a path is very small.
To simplify this expression, we shall require that $t_d\ll
D\sim5\times10^{12}$~sec---most of the neutrinos decay well before they
reach the earth. Otherwise, the flux is greatly reduced and the limits
are correspondingly weaker; the neutrinos may typically take a
path that causes them to decay well after our observations end. In
Sec.~\ref{sec:long} we discuss long lifetimes.

In this limit, the delta function becomes simply
$\delta\left(t-t_d(1-v\mu)\right)=(2\gamma
k/m)\delta(t_d-2\gamma kt/m)$, and we can perform the integral
over $t_d$:
\begin{equation}
  dN = {\Bg L_\#(E)\over4\pi D^2} f(E,k) {2k\over\mt}e^{-2kt/\mt}\,dt\,dE\,dk
\end{equation}
(Similar expressions have also been derived in \cite{ober,otherflux}.)
We shall assume that the neutrino-number luminosity is given by a
zero-chemical-potential Fermi-Dirac distribution with known temperature
and total neutrino energy, a reasonable
approximation\cite{burrowsannrev}. For now, we will present results for
low-mass neutrinos (\ie, $\mnu\ll \Tnu$),
\begin{equation}
  L_\#(E, \mnu=0) = {120\over7\pi^4} {E_T\over \Tnu^4} {E^2\over
    1 + e^{E/\Tnu}},
\end{equation}
where $E_T\simeq10^{53}$~erg is the total energy in one species of
massless neutrinos. We treat the case $\mnu\gtrsim\Tnu$ in
Sec.~\ref{sec:massive} below.

Finally, we can integrate the above expression over all neutrino
energies $E$ and over one time bin from $t$ to $t+\delta t$ to
get an expression for the spectrum of photons incident on the
detector during that time interval. For $\mnu\ll \Tnu$,
\begin{equation}\label{flux}
  \phi(k,t)= \int_{t}^{t+\delta t}
  {dN\over dk\;dt}dt = {\Bg\over4\pi D^2} {240\over7\pi^4} 
  {E_T\over \Tnu^2}h(k/\Tnu) e^{-2kt/\mt}\left(1-e^{-2k\delta t/\mt}\right).
\end{equation}
In this expression, the function $h(k/\Tnu)$ results from the
integral over the neutrino energies. It is of order unity for
the paramater ranges of interest, and it is largest in the case
of ``no flip,'' which we will assume from now on since it gives
the most conservative estimates of the parameters. In that case,
it is given by $h(y) = y\ln(1+e^{-y})$.

Although this signal depends nonlinearly on the parameter $\mt$,
the expression simplifies when $\mt$ is much greater or less
than $kt$ or $k\delta t$, where $k$ is a typical photon energy, 
giving
\begin{equation}\label{fluxcases}
  \phi(k,t) = 
  {\Bg\over4\pi D^2} {240\over7\pi^4} {E_T\over \Tnu^2}h(k/\Tnu)
  \times \cases{\delta(t),                        &
    \mbox{$\mt\ll k\delta t$} \cr 
                2k\delta t/\mt,           &$\mt\gg kt$ \cr}.
\end{equation}
In the former case of small $\mt$, there is no appreciable
relativistic delay before the decay of the neutrinos, so
essentially all of the daughter photons arrive in the first time
bin after the supernova.  In the case of large $\mt$, the flux
is essentially constant over the time of the observations, so
the signal is proportional to the width of the time bin.  Note
that only in the latter case does the fluence actually depend on
the value of $\mt$.

In order to calculate the expected signal from the theoretical
spectrum, we must fold that spectrum with the appropriate response
function. The signal expecied in the $i$th energy channel is
\begin{equation}
  S_i(t) = \int dk\,R_i(k) \phi(k,t) = \sum_j R_{ij} \phi_j(t)
\end{equation}
where $R_{ij}$ is the response of detector $i$ in energy bin
$k_j$, and $\phi_j(t)$ is the theoretical spectrum averaged over
energy bin $j$ at time (or time bin) $t$. This, combined with a
separate observation of the background rate in each of the four
detectors, provides the theoretical signal to be compared with
the observations.

\section{Gamma-Ray Data and Analysis}\label{sec:stat}
\begin{figure}
  \begin{center}\leavevmode\epsfysize=3in\epsffile{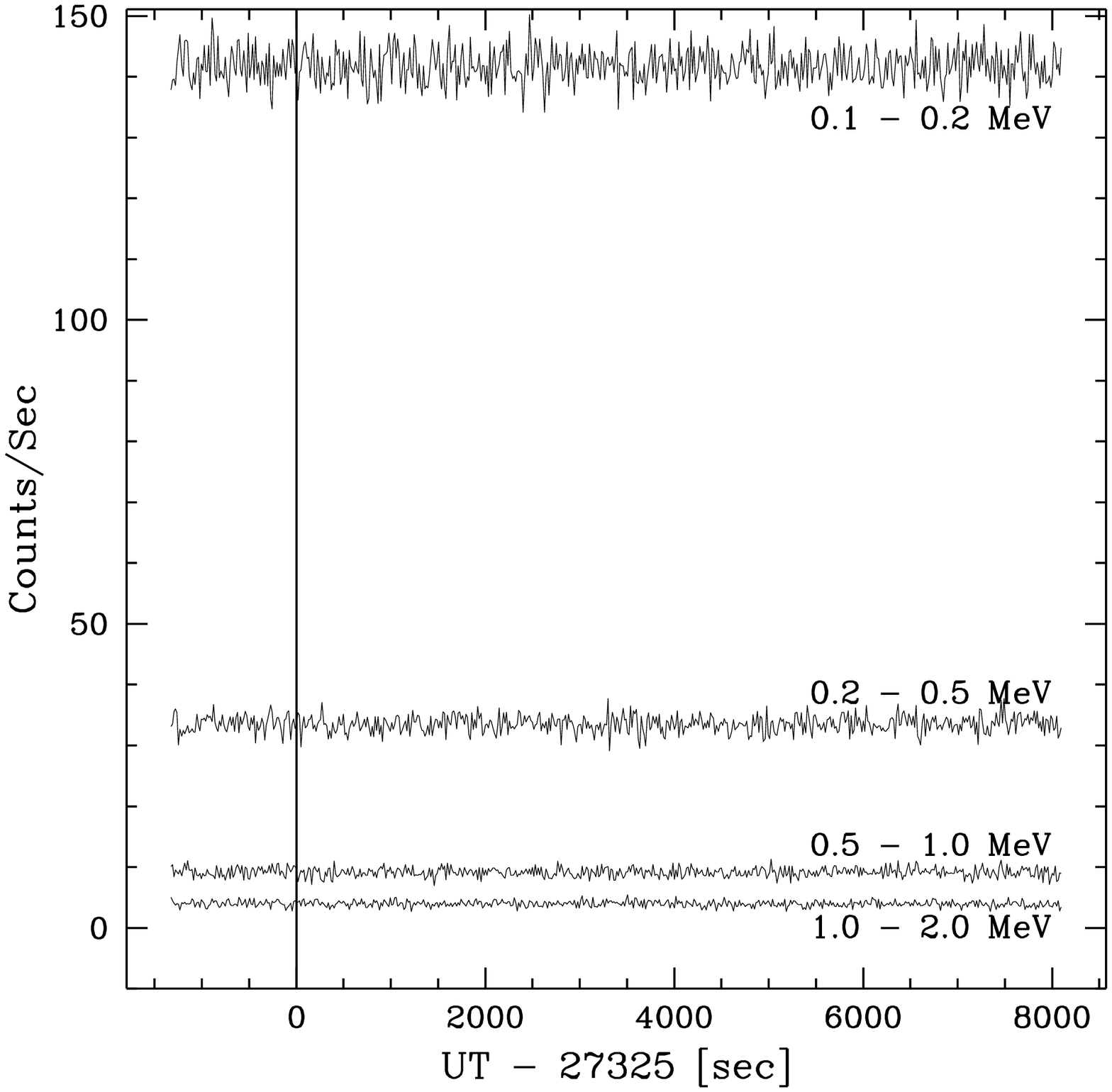}
  \end{center}
  \caption{The PVO GBD data for the time immediately before and after the
    arrival of the light from SN1987A at the PVO spacecraft (UT =
    27325). Time bins are either 12 or 16 sec; we show the average counts
    per second in each bin, for each energy channel, as marked.}
  \label{fig:data}
\end{figure}

To obtain our limits we use data from the supernova with data from
Pioneer Venus Orbiter Gamma Burst Detector (PVO GBD)\cite{pvo}, luckily
in operation in February, 1987.  The GBD has four energy channels,
roughly $100-200\keV$, $200-400\keV$, $500-1000\keV$, and $1-3\MeV$
towards the direction of the supernova, which was propitiously directly
overhead at the time, giving the maximum effective area.  We have data
for about 1500 sec prior to the supernova lights arrival at Venus (for
calculating the background), and for 8000 sec after, for time bins of
either 12 or 16 sec in duration.  We show the data in
Figure~\ref{fig:data}.  We have verified that there is no clear signal
in any of the four channels; further, the data is consistent with a
constant Poisson rate in each detector.

To calculate limits on our parameters $\Bg$ and $\mt$, we use the folded
spectrum $S_i(t)$ and our measurement of the background rate (observed
for a length of time $t_b$) in each detector to construct a likelihood
function given the observed gamma-ray counts in each detector. We assume
that the folded spectrum $S_i(t)$ gives the mean of a Poisson process
governing the detected number of counts, and that the rates for each
detector are high enough to approximate this by an appropriate Normal
distribution, for ease of calculation (in the $1-3\MeV$ bin, with the
lowest fluence, there are approximately 40 counts per bin).  This gives
\begin{equation}
  {\cal L}(\theta)=\prod N(D_{ij}; b_i\delta t_j+S_{ij}(\theta),\sigma^2_{ij})
\end{equation}
where
\begin{equation}
  N(x; \mu, \sigma^2) = {1\over\sqrt{2\pi\sigma^2}}
            \exp\left[-{1\over2}{(x-\mu)^2\over\sigma^2}\right]
\end{equation}
gives the Normal distribution, $b_i$ is the background rate in
detector $i$ (observed for a time $t_b$), $\delta t_j$ is the
length of time bin $j$, and $D_{ij}$, $S_{ij}$ are,
respectively, the observed and theoretical signal in those time
bins, where the latter is calculated with the set of parameters
represented by $\theta$.  Finally, the variance is given by
$\sigma_{ij}^2=S_{ij}+b_i\delta t_j(1+\delta t_j/t_b)$, the sum
of the theoretical variance of the signal and that due to the
background rate (including a small contribution reflecting the
uncertainty in that rate; this latter effect is somewhat more
difficult to include if a Poisson distribution is explicitly
used but is in any case negligeable). We define a $\chi^2$
statistic,
\begin{equation}
  \chi^2 \equiv -2\ln{\cal L} + {\rm const} = 
\sum_{ij}\left[\ln\sigma_{ij}^2 + {S_{ij}^2\over\sigma_{ij}^2} +
2{\left(b_i\delta t_j-D_{ij}\right)S_{ij}\over\sigma_{ij}^2} +
{\left(D_{ij}-b_i\delta t_j\right)^2\over\sigma_{ij}^2}
\right]
\end{equation}
Note that the model is nonlinear, and the variance
$\sigma^2_{ij}$ depends on the model parameters, so we have defined this 
quantity including the $\ln\sigma_{ij}$ term and that the usual
$\chi^2$ distribution does not apply; instead we must apply
Bayes' theorem and integrate over the likelihood to determine
confidence intervals on our parameters.

\begin{figure}
  \begin{center}\leavevmode\epsfysize=3in\epsffile{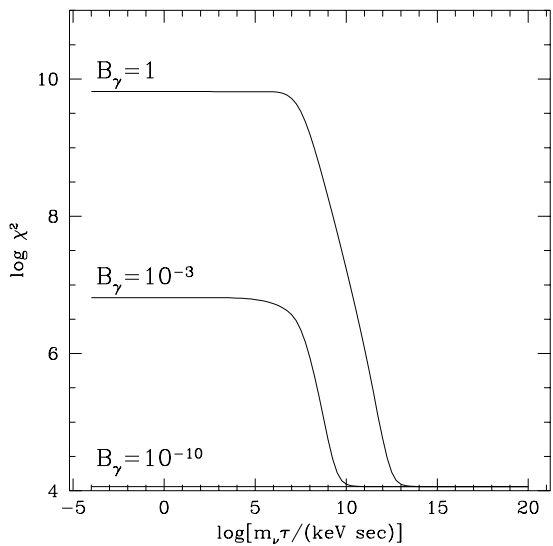}
  \end{center}
  \caption{The value of the $\chi^2$ statistic, defined in the text, as
    a function of the parameter $\mt$, for values of the branching ratio
    $\Bg$ as marked, for the decay process $\nu\rightarrow\nu'\gamma$.}
  \label{fig:chi2}
\end{figure}

Because of the two terms contributing to the variance, the form
of $\chi^2$ depends on which dominates. For $S_{ij}\gg b_i\delta
t_j$, $\sigma_{ij}^2\approx S_{ij}$, and the
$S_{ij}^2/\sigma_{ij}^2$ term dominates, so $\chi^2\sim \sum
S_{ij}$. When the neutrino signal is small, the background
contribution dominates, and $\chi^2\simeq{\rm const}$.
These regimes are shown in Figure~\ref{fig:chi2}, where we plot $\chi^2$ as a
function of $\mt$ for several values of $\Bg$. We have used a neutrino
temperature of 8~MeV, appropriate for mu and tau neutrinos.

Immediately, we see the character of the limits on the
parameters. For $\mt\lesssim 10^7\keV\sec$, $\chi^2\propto\Bg$;
in this regime only the data from the first time bin after the
supernova contributes.  Then $\chi^2\propto\Bg/\mt$ for
$\mt/\Bg\lesssim10^{13}\keV\sec$; now, the full data set
provides information.  Finally, $\chi^2={\rm const}$ for
$\mt/\Bg\gtrsim10^{13}\keV\sec$; in this regime the background
dominates over the theoretical signal. Note that this latter
area of parameter space provides the ``maximum likelihood'' (or
$\chi^2$ minimum); there is no neutrino signal and we calculate
only limits on parameters. In fact, there is a slight deficit of
counts with respect to the background calculated from the time
before the supernova; otherwise we might expect to see a weak
maximum likelihood somewhere in the large-$\mt$ regime.

\begin{figure}
  \begin{center}\leavevmode\epsfysize=3in\epsffile{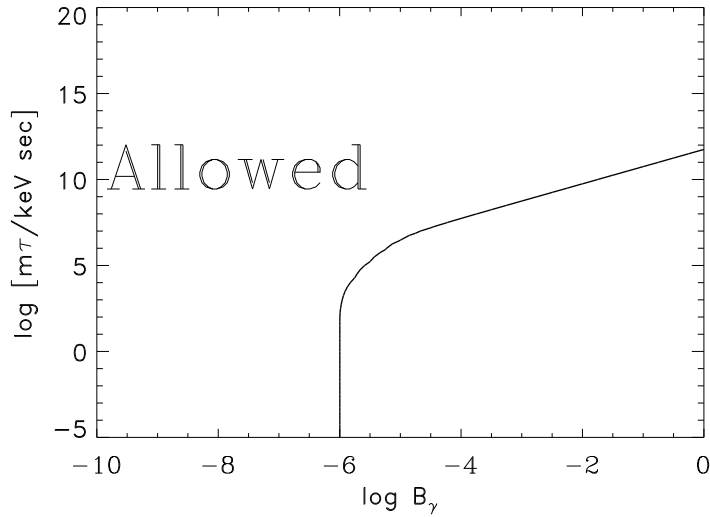}
  \end{center}
  \caption{Allowed region of the $\mt$-$\Bg$ plane, for the 2-body
    decay process $\nu\rightarrow\nu'\gamma$, corresponding to
    $\Delta\chi^2\le2357$ (see text). Here and below, contours continue
    to infinity as long as the appropriate assumptions, discussed in the
    text, still hold.}
  \label{fig:contour}
\end{figure}

In Figure~\ref{fig:contour}, we show a single contour of $\chi^2$ in the
$\mt$-$\Bg$ plain. From a ``Bayesian'' viewpoint, this figure
represents the the complete inference from the data; we
obviously would prefer to quote limits on the parameter space.

Because $\chi^2$ is constant for both large and small values of
the parameters $\mt$, we must be careful in normalizing our
probabilities. Without any other information to guide us, we
might expect the appropriate prior to use would be either
$p(\theta)\propto{\rm const}$ or $p(\theta)\propto1/\theta$
corresponding to constant probability per linear and logarithmic
interval, respectively; however, neither of these choices
converge when integrated out to infinity multiplied by a
constant likelihood. These priors, when normalized, assign a
vanishing weight to any finite region of parameter space;
usually, $\chi^2$ goes to infinity like the square of some
parameter in a linear gaussian model with independent errors, so
the wings are automatically disfavored regardless of the prior.

To get meaningful results in this case we must make sure that the
interesting regime of $-6\lesssim\log\Bg < 0$,
$5\lesssim\log(\mt/\keV\sec)\lesssim13$ is given finite prior weight.
This corresponds to choosing a prior with some cutoff outside of this
region, and will result in a limit corresponding to a value of $\chi^2$
somewhere along the slope between its small and large values. To connect
with other analyses, we will adopt the following ``frequentist''
procedure. We will assume that $\chi^2$ is distributed in a $\chi^2$
distribution considering the data as a frequentist random variable. We
have 550 time bins and four detectors, so there are 2200 degrees of
freedom. For this distribution, a one-sigma fluctuation corresponds to
$\Delta\chi^2=2230$, a three-sigma (or 99\%) fluctuation to
$\Delta\chi^2=2357$; we choose the latter as our limit; from the shape
of the likelihood function it is clear that any comparable
$\Delta\chi^2$ will give similar bounds. We also note that a signal in
the small-$\mt$ regime may not be detectable with this algorithm; the
absence of a local minimum in that region, however, implies that this
should not be a significant worry. (We have also used a $\chi^2$ defined
rigorously from the true Poisson distribution, which effects the limits
on the parameters by somewhat under one order of magnitude throughout.)
The allowed region is shown in Figure~\ref{fig:contour}. It corresponds
to
\begin{eqnarray}  \label{bound1}
  \Bg&<&3\times 10^{-7} \phantom{\mt\over \keV\sec }\qquad 
         \mt\lesssim 10^{6} \keV\sec \\
  \Bg&<&2\times 10^{-13} {\mt\over \keV\sec} \qquad \mt\gtrsim 10^{6} \keV\sec.
\end{eqnarray}
for neutrinos with a temperature of 8~MeV. (The limits scale roughly as
$T^{-2}$.) The latter bound corresponds to $\mt_\gamma > 5\times10^{12}
\keV\sec$.  This is less restrictive than the
limits of Oberauer et al.\cite{ober}, due to the fact that the PVO GBD
could only detect gamma rays with energies below $3\MeV$, compared to
$25\MeV$ for the SMM satellite. However, we believe this analysis to be
more rigorous and the PVO data to be of higher quality.

\section{Other Decay Modes}

So far we have considered only photons produced from the
simplest radiative decay of a massive neutrino species:
$\nu\rightarrow\nu'\gamma$, but many other channels are
possible.  For these reactions, where the rest-frame photon
energy is no longer given by the simple ${\bar k} = \mnu/2$, we
must allow for a distribution of decay products: $f({\bar k},
{\bar\mu})d{\bar k} d{\bar\mu}$ gives the fraction of photons
produced with rest-frame energy ${\bar k}$ into angle
${\bar\mu}=\cos{\bar\theta}$. Then, the final spectrum is given
by
\begin{equation}
{dN\over dk dt} = {1\over4\pi D^2} {\Bg\over\tau}\int dE\; L_\#(E) 
\int d{\bar\mu}\;
f\left[{k\over\gamma(1+v{\bar\mu})},{\bar\mu}\right]
e^{-\gamma(1+v{\bar\mu})t/\tau}
\end{equation}
where we have still assumed that the decays occur near the
supernova, and the integral is taken from $k$ to $\infty$. (We
discuss the case of large masses, $\mnu\gtrsim\Tnu$ in
Sec.~\ref{sec:massive} below.)  For the appropriate choice of
$f({\bar k}, {\bar\mu})\propto\delta({\bar k}
-\mnu/2)f_1({\bar\mu})$, we reproduce the earlier 2-body decay
formula, Eq.~(\ref{flux}).

In particular, we consider the \brem process,
$\nu\rightarrow\nu'e^+e^-\gamma$, where typically $\nu=\nu_\tau$,
$\nu'=\nu_e$.  Because this is no longer a two-body decay, the spectrum
in angle and energy of the daughter photons is considerably more
complicated, and there has been no exact calculation performed.
Following Oberauer et al. \cite{ober}, we make several simplifying
assumptions. First, we assume isotropy of the photons in the rest frame
of the neutrinos; this is reasonable if the helicity states of the
parent neutrinos are produced in equal numbers. This gives $f({\bar k},
{\bar\mu})=f({\bar k})/2$ (which still implicitly depends on ${\bar\mu}$
through the Lorentz transformation to the lab frame).

Up to factors of order unity, we assume after Oberauer et al.\ 
that the \brem energy spectrum is given by
\begin{equation}\label{bremspec}
  {d\Gamma_{\rm br}\over dk} 
  \simeq
  {\alpha\over\pi}{\Gamma_0\over k}={\alpha\over\pi}{1\over k\tau_e}
\end{equation}
where $\Gamma_0$ and $\tau_e$ refer to the process without a daughter
photon: $\nu\rightarrow\nu'e^+e^-$, allowing us to absorb a branching
ratio factor into $\tau_e=\tau/B_e$. Now, we can write the gamma-ray
flux as
\begin{eqnarray}
  {dN\over dk dt}&=&
  {1\over4\pi D^2} {1\over\mnu\tau_e} {\alpha\over\pi}
  \int dE\; L_\#(E) {E\over k} \int_{-1}^{+1} {d{\bar\mu}\over2}\;
  \left(1 + v{\bar\mu}\right)e^{-\gamma(1+v{\bar\mu})t/\tau}\nonumber\\
  &=& {1\over4\pi D^2} {1\over\mnu\tau_e} {\alpha\over\pi}
  \int dE\; L_\#(E) {E\over k} \left.
  {1\over2v}\left(\mt\over Et\right)^2
  e^u(u-1)\right|_{u=-(1-v)Et/\mt}^{-(1+v)Et/\mt}
\end{eqnarray}
If we assume that $\gamma t\ll \tau$, the angular average
becomes unity. (A Maxwell-Boltzmann distribution for $L_\#(E)$
then reproduces the expression of Eqs.  (6-7) of Oberauer et
al.\cite{ober})  With the additional assumption that a negligible fraction
of the decays occur inside of the progenitor and that detected
gamma rays have energies $k\ll T$ (\ie, $E_{\rm min}\ll T$), we
get the simple formula
\begin{equation}\label{brem}
{dN\over dk dt}= {E_T\over4\pi D^2} {1\over\mnu\tau_e}
{\alpha\over\pi} {1\over k}
\end{equation}
where $E_T$ again gives the total energy in the decaying
neutrino species. Note that this expression is correct for any
form of $L_\#(E)$. At early times the flux is constant, although
there will be an exponential decay for $\gamma t\gtrsim\tau$. In
the two-body case, the timescale for gamma-ray detection is set
by $\mt/k$; in the \brem case, by $\mt/E\sim\mt/T$.

Assuming that most neutrinos will have $E\sim T$, the
assumptions we have made require that $t/\tau\ll\mnu/T \ll 1$
for this formula to hold. Note that this is a restriction on the
parameter space of $\mnu$ and $\tau$, since we have
approximately 8000 sec of data, and the temperature is of order
8 MeV for any neutrino species.

\begin{figure}
  \begin{center}\leavevmode\epsfysize=3in\epsffile{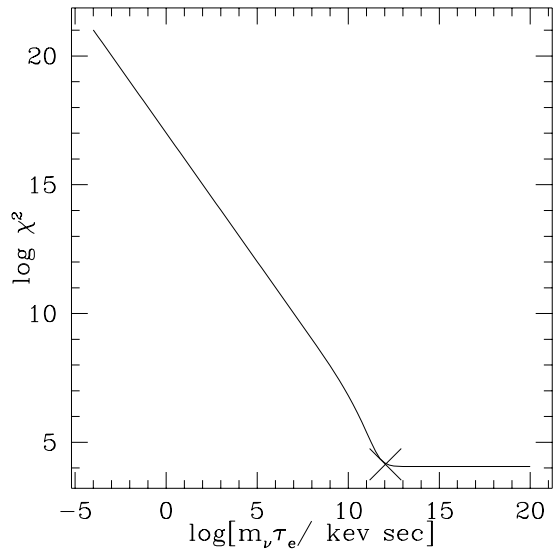}
  \end{center}
  \caption{The value of the $\chi^2$ statistic, defined in the text, as
    a function of the parameter $\mt_e=\mt/B_e$, for the \brem
    process. Also shown is the location of the $\Delta\chi^2\le2357$ limit.}
  \label{fig:chi2-brem}\end{figure}

Because the flux is proportional to $1/\mnu\tau_e$, we will be
able to put limits on the combination $\mnu\tau_e=\mt/B_e$.
Moreover, because the dependence is the same as the large-$\mt$
regime of the 2-body decay case (cf.  Eq.~(\ref{fluxcases})),
the probability density will be of the same form. When the
variance is dominated by the signal, $\chi^2\sim\sum
S_{ij}\propto1/\mt_e$; when it is dominated by the background,
$\chi^2\sim{\rm const}$ (the same constant as in the 2-body
case). The crossover occurs at $\mt_e\approx10^{15} \keV\sec$.
The value of $\chi^2$ for this model is shown in
Figure~\ref{fig:chi2-brem}.  Using the same quasi-frequentist
definition of a 99\% confidence level gives limits of
\begin{equation}
  \mnu\tau_e>1.5\times 10^{12}\keV\sec \qquad{\rm or}\qquad
  {B_e\over\mt}<7\times 10^{-13}\keV^{-1}\sec^{-1}.
\end{equation}
Because the \brem spectrum peaks at a lower energy, and
due to the long time baseline of the PVO data, this limit is
comparable to other SN1987A limits for this decay
channel\cite{smm1,ober}, and we believe more reliable, due to the
higher-quality data set.

\section{Very Massive Neutrinos}\label{sec:massive}
All of these expressions are considerable more complicated in
the case $m\gtrsim T$. We will still assume a
zero-chemical-potential FD distribution, this time applying to
massive particles:
\begin{equation}
L_\#(E, \mnu) = {120\over7\pi^4} {E_T\over \Tnu^4}
j\left(\mnu/\Tnu\right){E\sqrt{E^2+\mnu^2}\over1 + e^{E/\Tnu}},
\end{equation}
with a ``suppression factor,'' $j(x)$, along with the requirement that $E
> \mnu$. The factor $j(x)$ is just the usual Boltzmann suppression
($j(x)\propto x^{3/2}e^{-x}$, for $x\gg1$); we use an approximation that
is good for $m/T\lesssim{\rm few}$, $j(x)\simeq \exp\left(-0.15
  x^2\right)$. (Sigl \& Turner\cite{sigl} have calculated the effect of
the changing neutrinosphere temperature and radius on this naive
expectation; the effect is small for $\mnu\lesssim40\MeV$, at least for
$\tau\gtrsim10^{-2}\sec$.)  For low-mass neutrinos we assumed $k>\mnu$;
now, we can only integrate over neutrino energies greater than
$\max(\mnu, k)$.  In this expression, $E_T\simeq10^{53}$~erg remains the
total energy for the low-mass case; the total energy released is
$E_Tj(\mnu/\Tnu)$; which for large masses is less than $10^{53}$~erg
since $j<1$

In addition to the mass threshold effects, with massive
neutrinos ($\mnu\gtrsim k$) we must now take into account the
loss of any photons produced inside the envelope of
the supernova, $R_{\rm env}=100c\sec$. Thus, we require that
$vt_d>R_{\rm env}$, or 
\begin{equation}
  E>E_{\rm env} = \mnu\sqrt{1+\left(R_{\rm env} \mnu\over 2kt\right)}.
\end{equation}
Note that $E_{\rm env}>\mnu$, so this supersedes the requirement
that $E>\mnu$, but the requirement that $E>k$ remains. Thus, we
must integrate over neutrino energies from $E_{\rm
  min}=\max(k,E_{\rm env})$. This integral, the equivalent of
$h(k/T)$ above, cannot be done in closed form, but again it can
be approximated by a gaussian (at least for the isotropic case
$f(E,k)=1/p$):
\begin{equation}
  {dN\over dk\;dt} = {\Bg\over4\pi D^2}
  j\left(\mnu/\Tnu\right) {120\over7\pi^4}
    {E_T\over \Tnu^2} {2k\over\mt}e^{-2kt/\mt}g(E_{\rm min}/T)
\end{equation}
with
\begin{equation}
  g(x) = \int_x^\infty dy\; {x\over1+e^x} \simeq {\pi^2\over12}e^{-0.2x^2};
\end{equation}
the factor of $0.2$ in the exponent approximates the shape
of the integral for $x\lesssim{\rm few}$.  This differs from the
massless case by a total suppression factor
\begin{equation}
  {j\left(\mnu/\Tnu\right)
  g\left(E_{\rm min}/ T\right)\over h(k/T)}\simeq
\exp\left[-0.15 \left(\mnu\over\Tnu\right)^2
        - 0.2 \left(E_{\rm min}\over \Tnu\right)^2\right]
\end{equation}
Since $E_{\rm min}\ge\mnu$, this is always less than
$\exp[-0.35(\mnu/\Tnu)^2]$ for interesting masses
$\mnu\gtrsim\Tnu$; unfortunately, the time now appears in the
expression for $E_{\rm min}$, so the $dt$ integral is no longer
trivial. First, then, let us consider the suppression factor if
we ignore the effect of decays inside the supernova envelope,
integrating from $E_{\rm min} = \max(\mnu, k)$. Then the
time integral can be done as in the low-mass case, and we can
simply write down the time-independent suppression factor
$s = j(\mnu/T) g(E_{\rm min}/T)\simeq j(\mnu/T)g(\mnu/T)$ (if we first
set $h(k/T)=1$ when numerically calculating the limits as above).

These mass effects enable us to break the degeneracy between $\mnu$ and
$\tau$, at the price of requiring the more complicated analysis of a
three-dimensional parameter space.  To simplify matters, we will base
the results for massive neutrinos directly on the limits from the
low-mass case. That is, we will calculate the limits as before, and then
apply the suppression factor at the end. We can do this because the
suppression factor comes into the expression for the flux in exactly the
same way as the branching ratio $\Bg$, so we translate limits on $\Bg$
in the massless case to limits on $\Bg\times s$, where $s$ is the
$k$-independent part of the suppression factor. In addition, we do the
calculation for an isotropic decay, and assuming $k<\mnu$. For two-body
decay, this results in the limit
\begin{eqnarray}
  s\Bg&<&3\times 10^{-7} \phantom{\mt\over \keV\sec }\qquad 
         \mt\lesssim 10^{6} \keV\sec \\
  s\Bg&<&6\times 10^{-14} {\mt\over \keV\sec} \qquad \mt\gtrsim 10^{6} \keV\sec.
\end{eqnarray}

\begin{figure}
  \begin{center}\leavevmode\epsfysize=3in\epsffile{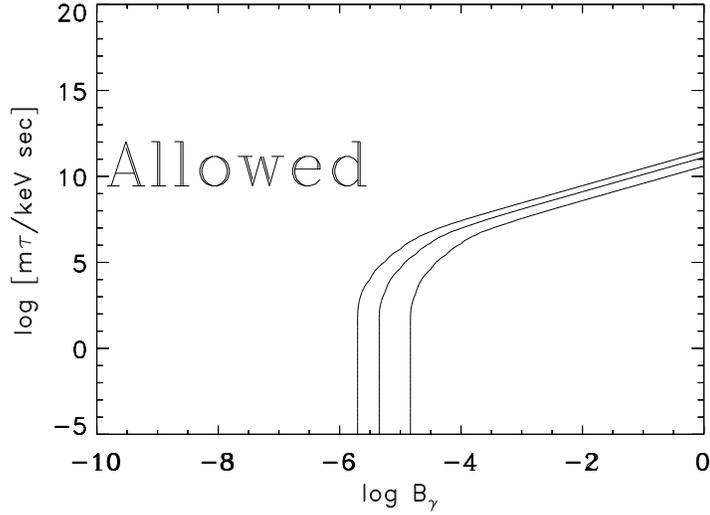}
  \end{center}
    \caption{Allowed region of the $\mt$-$\Bg$ plane, for 
      neutrinos of mass 20~MeV, 30~MeV, and 40~MeV (from left to right),
      for the 2-body decay process, corresponding to
      $\Delta\chi^2\le2357$ (see text).}
  \label{fig:30MeV}
\end{figure}

If we allow the effect of decays inside the progenitor envelope,
the calculation is somewhat more complicated, and the integral
over each time bin can no longer be done in closed form. We must
now recompute everything at each pair of $\mnu$ and $\tau$. We
show the results of such a calculation for several values of the
neutrino mass in
Figure~\ref{fig:30MeV}; the limits are not too different from
those with the simpler time-independent suppression.

For the \brem process, the suppression of a high-mass neutrino flux is
simpler to calculate because the required integral is simply $\int dE\;E
L_\#(E)$, the total energy in the massive neutrino species (cf.
Eq.~\ref{brem}; we have again assumed $k\ll T$, so the initial integral
over $\mu$ simplifies). Again, we integrate from the same $E_{\rm min} =
\max(k, E_{\rm env})$; the suppression is given by the Boltzmann factor
$j(E_{\rm min}/T)$. For much of parameter space, this is simply the
expected $j(\mnu/T)$. Again, the time-dependence of $E_{\rm env}$ does
not change the limits significantly. The allowed parameter space for the
\brem process for a mass of 30 MeV is
\begin{equation}
  {\mnu\tau_e\over j(\mnu/T)}>1.5\times 10^{12}\keV\sec \qquad{\rm or}\qquad
  j(\mnu/T){B_e\over\mt}<7\times 10^{-13}\keV^{-1}\sec^{-1}.
\end{equation}

\section{Nearly Degenerate Neutrinos}\label{sec:degen}
Thus far, we have assumed that the daughter neutrino in the
$\nu\to\nu'\gamma$ channel is much less massive than the parent
neutrino. If, however, the mass of the daughter is appreciable, the
energy of the photon will be decreased by a factor $\delta
m^2/m^2\equiv(m_1^2-m_2^2)/m_1^2$. This may improve our limits: It would
shift the bulk of the photons down from energies too high to detect into
one or more of the energy channels of the PVO detector. For massless
daughter neutrinos, of order 1/10 of the photons can be detected;
therefore we might expect limits as much as an order of magnitude
stronger. In fact, for this case, the lower energy window of the PVO
detector, down to $0.1$ MeV, compared with the SMM window, sensitive
only above $4.1$ MeV, is actually an advantage.

To make the matter more precise, we see that in the case of nearly
degenerate neutrinos, we make the change
\begin{equation}
  f(E, k) dk\rightarrow f[E, (m^2/\delta m^2)k] (m^2/\delta m^2) dk
\end{equation}
where we now are contrained to have photon energies $k<(\delta m^2/m)E$.
This, in turn, results in changing $h(k/T)\to(m^2/\delta m^2)
h[(m^2/\delta m^2)k/T]\approx (m^2/\delta m^2) h(k/T)$, up to
a constant of order one. As expected, the flux is enhanced
by a factor of $(m^2/\delta m^2)$ at most energies, so we can make the
substitution $\Bg\to\Bg(m^2/\delta m^2)$ in our limits.

\begin{figure}
  \begin{center}\leavevmode\epsfysize=3in\epsffile{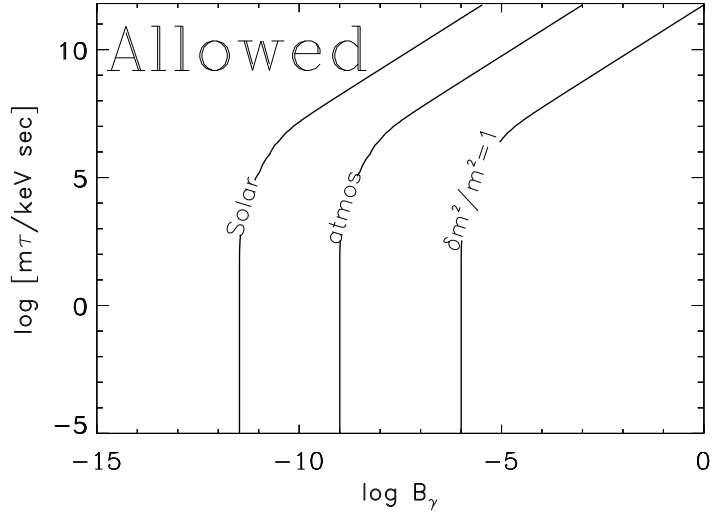}
  \end{center}
    \caption{Allowed region of the $\mt$-$\Bg$ plane, for nearly
      degenerate neutrinos, with $\delta m^2/m^2=10^{-3}$
      (``atmospheric''), $3\times10^{-6}$ (``solar''), as well as
      $\delta m^2/m^2=1$,
      as labelled.}
  \label{fig:degen}
\end{figure}

In particular, we are interested in two regimes of $\delta m^2/m^2$.  To
account for the recent evidence of a neutrino mass eigenstate, we set
$m=3{\rm\ eV}$\cite{lsnd}. The most attractive solution to the solar
neutrino problem requires $\delta m^2=3\times10^{-5}{\rm\ eV}^2$, so
$\delta m^2/m^2=3\times10^{-6}$\cite{solarnu}. To account for the
atmospheric $\nu_\mu$ deficit, $\delta m^2=1\times10^{-2}{\rm eV}^2$, so
$\delta m^2/m^2=1\times10^{-3}$\cite{atmosnu}.  Roughly, we see that our
limits will become stronger by factors order $3\times10^5$ and $10^3$,
respectively. In Figure~\ref{fig:degen}, we show the limits on the
neutrino parameters for these two cases.

\section{Long Lifetimes}\label{sec:long}
For long lifetimes (such that the average decay time of the neutrino is
comparable to or longer than the travel time to the detector), the above
formalism becomes too cumbersome, because we must integrate over a
complicated set of possible paths for the neutrino and daughter
photon. In this case, we will make several simplifications.
At first, we will only concern
ourselves with the total gamma-ray fluence from the decays, integrated
over time. Then, we will integrate over the decay time, $0\le t_d\le D$:
\begin{equation}
  {dN\over dk} =  {\Bg \over4\pi D^2} \int dE\; L_\#(E) f(E,k)
\left[1-e^{-D\mnu/E\tau}\right]
\end{equation}
where, as before, $f(E,k)$ gives the fraction of neutrinos with energy
$E$ decaying into photons with energy $k$. This expression is to be
compared with those presented in \cite{smm1}.  The cost of the
simplicity of this expression is the inability to determine the exact
time of a photon's arrival. For neutrinos and photons travelling on a
straight path ($\mu=1$, appropriate for relativistic particles), the
arrival time after the supernova light-pulse is
$t=t_d(1-v)\simeq(t_d/2)\mnu^2/E^2$. For long lifetimes, we will be
concerned with neutrinos that decay late in their flight: $t_d\sim D$.
Using this as a typical $e$-folding time, we have the ansatz that
$
  (dN/dt) \propto  \exp\left(-2t E^2/D\mnu^2\right).
$
For $t\gtrsim(D/2)\mnu^2/E^2$, this should express the character of the
time-dependence. Two effects are explicitly missing from this formula:
the extra time delay from non-straight paths (of the same order as the
delay already considered) and the photon energy dependence of the
time-delay. Moreover, the time-dependence will not have exactly this
shape; for short times it does not contain the expected slow
rise from zero flux, so it is probably safest to use this formula
integrated over the entire duration of the experiment, and not rely on
the detailed time evolution, leaving us finally with
\begin{equation}
  \left.{dN\over dk}\right|_{\delta t} \simeq  
  {\Bg \over4\pi D^2} \int dE\; L_\#(E) f(E,k)\; 
  \left(1-e^{-2\delta t E^2/D\mnu^2}\right)
  \left(1-e^{-D\mnu/E\tau}\right).
\end{equation}
Unfortunately, the integration over neutrino energy $E$ is considerably
more complicated than before, but we can approximate the two
exponential decays for various regimes:
\begin{eqnarray}\label{taucases}
  \left(1-e^{-2\delta t E^2/D\mnu^2}\right)&&
  \left(1-e^{-D\mnu/E\tau}\right)\\
  &&\approx\left\{
    \begin{array}{lll}
      1; & m\lesssim E\sqrt{2\delta t/D}, & \mnu/\tau\gtrsim E/D \nonumber\\
      {2\delta t E^2/ D\mnu^2};
        & m\gtrsim E\sqrt{2\delta t/D},  & \mnu/\tau\gtrsim E/D \nonumber\\
      {D\mnu/ E\tau};
        & m\lesssim E\sqrt{2\delta t/D}, & \mnu/\tau\lesssim E/D \nonumber\\
      {2\delta t E/\mt};
        & m\gtrsim E\sqrt{2\delta t/D},  & \mnu/\tau\lesssim E/D.
    \end{array}\right.
\end{eqnarray}
Numerically, these breaks occur at
\begin{equation}
   m\simeq E\sqrt{2\delta t/D}\simeq 680 {\rm\ eV}
   {E\over12{\rm\ MeV}}\left(\delta t\over8500{\rm\ sec}\right)^{1/2}
;\qquad
   \mnu/\tau\simeq E/D\simeq 2\times10^{-6} {\rm\ eV/sec}{E\over12{\rm\ Mev}}.
\end{equation}
where $E=12$ MeV is a typical energy for a $T_\nu=8$ MeV blackbody. In
terms of the lifetime $\tau$, the latter limit occurs at $\tau\simeq
D\mnu/E\approx5\times10^5{\rm\ sec}\;(\mnu/{\rm eV})$---for masses
$\mnu\sim1$\ MeV, this is roughly $\tau\sim D$. When this is
proportional to $\delta t$, the flux is approximately constant;
otherwise, the entire pulse is detected (and its shape is irrelevent).
Putting all of this together, and doing the integral over $E$ gives
\begin{equation}\renewcommand{\arraystretch}{2.0}
  \left.{dN\over dk}\right|_{\delta t} \approx {\Bg\over4\pi D^2} {120\over7\pi^4}
  \times\left\{\begin{array}{l}
\displaystyle {E_T\over T_\nu^2} h_0(k/T) \\
\displaystyle 2{E_T\over \mnu^2}{\delta t\over D} h_2(k/T) \\
\displaystyle {E_T \mnu\over T_\nu^3}{D\over \tau} h_{-1}(k/T) \\
\displaystyle 2{ E_T\over T_\nu\mnu}{\delta t\over\tau} h_1(k/T)
  \end{array}\right.
\end{equation}
where $h_n(y)=\int_y^\infty x^{n+1}/(1+\exp{x})$ is similar to $h(y)$
above, and the four cases correspond to those in Eq.\
(\ref{taucases}). Here, we have assumed an isotropic distribution of
decays in the rest frame. As before, these expressions hold for
$\mnu\lesssim\Tnu$ and must be modified with the appropriate suppression
factor otherwise.

Now, we can just put these results through our statistical machinery and
find limits on the parameters. We will write the flux as
\begin{equation}
  \left.{dN\over dk}\right|_{\delta t} \approx 
 {1\over4\pi D^2} {120\over7\pi^4} {E_T\over T_\nu^2} h(k/T) \times \Bg A(\mnu,\tau)
\end{equation}
where $A$ is the appropriate dimensionless combination of $\mnu$ and
$\tau$, along with $D$, $T_\nu$ and $\delta t$; the data give us limits
on $A$ in each $(\mnu,\tau)$ regime (assuming that we can write our
prior information as a simple probability distribution for $A$). This
gives an approximate 99\% confidence limit of $\Bg
A\lesssim1\times10^{-6}$ for $T_\nu=8\MeV$ or,
\begin{eqnarray}\label{taulim}
  \Bg\lesssim 1\times10^{-6} \left(T_\nu\over8\MeV\right) &\qquad&
  \mnu\lesssim 0.4\keV, \mnu/\tau\gtrsim 1.2\times10^{-9} {\keV/\sec};
\nonumber\\
  \mnu\gtrsim 155\keV \Bg^{1/2} &\qquad&
  \mnu\gtrsim 0.4\keV, \mnu/\tau\lesssim 1.2\times10^{-9} {\keV/\sec};
\nonumber\\
  {\Bg\mnu\over\tau}\lesssim 1.4\times10^{-15}\keV\sec^{-1}
  \left(T_\nu\over8\MeV\right)^3 &\qquad&
  \mnu\lesssim0.4\keV, \mnu/\tau\gtrsim 1.2\times10^{-9} {\keV/\sec};
\nonumber\\
  \mnu\tau\gtrsim 1.4\times10^{14} \keV\sec\Bg
  \left(T_\nu\over8\MeV\right)^{-1} &\qquad&
  \mnu\gtrsim0.4\keV, \mnu/\tau\lesssim 1.2\times10^{-9} {\keV/\sec}.
\end{eqnarray}
Where the regimes overlap, these limits are comparable to those
calculated with the more detailed models above---because we can only
calculate limits on parameters, the details of the data and the analysis
are unimportant (in fact, the limits of Eq.~(\ref{taulim}) are {\em
  stronger}\/ than, for example, Eq.~(\ref{bound1}) above; the earlier, more
detailed calculation is probably the more appropriate limit).
Again, for neutrinos with $\mnu\gtrsim T_\nu$, these limits are modified
with $\Bg\to s\Bg$.

For the \brem channel, the flux is changed due to the different
kinematics of the decay ({\it i.e.}, the rest frame spectrum of
Eq.~(\ref{bremspec})):
\begin{equation}
\left.{dN\over dk}\right|_{\mbox{brem}} = 
{\alpha\over\pi}{2T_\nu^2\over k\mnu}\times
\left. dN\over dk\right|_{\mbox{2-body}}
\end{equation}
(in addition, the functions $h_n$ should also be modified to
$h_{n+2}$). This is a significant increase in flux at for $k\mnu\lesssim T^2$.
As before, we see that the \brem spectrum at the detector is
proportional to $1/k$. Now, the limits correspond to $\Bg A
T/m\lesssim3\times10^{-5}$ or
\begin{eqnarray}\label{taulim-brem}
  \mnu\gtrsim 2.7\times10^8\keV\Bg \left(T_\nu\over8\MeV\right) &\qquad&
  \mnu\lesssim 0.4\keV, \mnu/\tau\gtrsim 1.2\times10^{-9} {\keV/\sec};
\nonumber\\
  \mnu\gtrsim 800\keV \Bg^{1/3} \left(T_\nu\over8\MeV\right)&\qquad&
  \mnu\gtrsim 0.4\keV, \mnu/\tau\lesssim 1.2\times10^{-9} {\keV/\sec};
\nonumber\\
  \tau/\Bg=\tau_e\gtrsim 1.9\times10^{16}\sec &\qquad&
  \mnu\lesssim0.4\keV, \mnu/\tau\gtrsim 1.2\times10^{-9} {\keV/\sec};
\nonumber\\
  \mnu\gtrsim 1.4\times10^{8} \keV\left(\tau/\Bg\over\sec\right)^{-1/2}
  \left(T_\nu\over8\MeV\right) &\qquad&
  \mnu\gtrsim0.4\keV, \mnu/\tau\lesssim 1.2\times10^{-9} {\keV/\sec}.
\end{eqnarray}

\section{Discussion}
SN1987A not only confirmed astrophysicists' standard model of the Type
II (core collapse) supernovae, but also provided a laboratory for
studying the properties of neutrinos. The fluence of neutrinos at earth
was approximately $10^{11}{\rm cm}^{-2}$ per species. This large fluence
and the space-borne gamma-ray detectors operating on SMM and PVO have
allowed stringent limits to be placed on the radiative decay of neutrinos.

 Although there is only 232 seconds of data, in
a single time bin, the SMM detectors were sensitive up to energies of 25 MeV, and so
would have more likely to detect gamma rays from decaying SN1987A
neutrinos with a temperature of 4-8 MeV. Because the limits are
determined by the region of parameter where the background becomes
comparable to the signal (see the discussion in Sec.~\ref{sec:stat}
above), the long time base and greater resolution is actually of little
use in determining the limits on the parameters. To show this, we have
performed our analysis with the SMM data as presented in \cite{ober},
using crude estimates of the response matrix and effective area of the
detector. As expected, the results are comparable to, although slightly
weaker than, those quoted in
that reference. However, in the case where the gamma rays are produced
by \brem or the neutrino mass states are nearly degenerate, the PVO
limits are comparable or more stringent. Finally, the amount of and
quality of the PVO data adds additional confidence to the SMM-based limits.


\acknowledgements Portions of this work was performed under NASA Grant
NAG~2-765. In addition, MST is supported by the DOE (at Chicago) and
NASA (at Fermilab). We would like to thank the PVO GBD team, and
especially Ed Fenimore, for access to and expertise in the analysis of
the PVO data, and the response matrix for the PVO GBD. MST
thanks F.~Vanucci for arousing our interest in the possibility of the
neutrinos with nearly degenerate mass eigenstates.  AJ would like to
especially thank Carlo Graziani and Peter Freeman for their patient
explanations of the theory and practice of statistics.


\end{document}